\documentstyle[mncite,psfig]{mn}

\title[Roche tomography of CVs: I. artefacts and techniques]
{Roche tomography of cataclysmic variables: I. artefacts and techniques}

\author[C.\,A.\ Watson and V.\,S.\ Dhillon]
{C.\,A.\ Watson and V.\,S.\ Dhillon\\
Department of Physics and Astronomy, University of Sheffield, 
Sheffield S3 7RH, UK \\}

\date{\center{\Large Accepted for publication in the Monthly 
Notices of the Royal Astronomical Society \\ 
\vspace{.5cm} 23/01/2001}} 

\begin{document}
\maketitle

\begin{abstract}
Roche tomography is a technique used for imaging the Roche-lobe filling
secondary stars in cataclysmic variables (CVs). In order to interpret
Roche tomograms correctly, one must determine whether features in the
reconstruction are real, or due to statistical or systematic errors. 
We explore the effects of systematic errors using reconstructions of simulated
datasets and show that systematic errors result in characteristic
distortions of the final reconstructions that can be identified
and corrected. In addition, we present a new method of estimating statistical
errors on tomographic reconstructions using a Monte-Carlo bootstrapping 
algorithm and show this method to be much more reliable than Monte-Carlo
methods which `jiggle' the data points in accordance with the size of their
error bars.

\end{abstract} 


\section{Introduction}
\label{sec:introduction}

The secondary, Roche-lobe filling stars in CVs are
key to our understanding of the origin, evolution and behaviour of this
class of interacting binary. To best study the secondary stars in CVs we
would ideally like direct images of the stellar surface. This is
currently impossible, however, as typical CV secondary stars have radii of 
400 000 km and distances of 200 parsecs, which means that to detect a feature
covering 20 per cent of the star's surface requires a resolution of 
approximately 1 micro-arcsecond, 10 000 times greater than the diffraction 
limited resolution of the world's largest telescopes. \scite{rutten94}
described a way around this problem using an indirect imaging
technique called {\em Roche tomography} which uses phase-resolved spectra
to reconstruct the line intensity distribution on the surface of the secondary
star.

Obtaining surface images of the secondary star in CVs
has far-reaching implications.
For example, a knowledge of the irradiation
pattern on the inner hemisphere of the secondary star in CVs is
essential if one is to calculate stellar masses accurately enough to
test binary star evolution models (see \pcite{smith98}). Furthermore,
the irradiation pattern provides information on the geometry of the
accreting structures around  the white dwarf (see \pcite{smith95}).
Perhaps even more importantly, surface images of CV secondaries can be
used to study the solar-stellar connection.  It is well known that
magnetic activity in isolated lower-main sequence stars increases with
decreasing  rotation period (e.g.~\pcite{rutten87}). The most rapidly
rotating isolated stars of this type have rotation periods of $\sim8$
hours, much slower  than the synchronously rotating secondary stars
found in most CVs. One would therefore expect CVs to show even higher
levels of magnetic activity. There  is a great deal of indirect
evidence for magnetic activity in CVs -- magnetic activity cycles have
been invoked to explain variations in the orbital periods, mean
brightnesses and mean intervals between outbursts in CVs
(see \pcite{warner95a}). The magnetic field of the secondary star is
also believed to play a crucial role in angular momentum loss via
magnetic braking in longer-period CVs, enabling CVs to transfer mass
and evolve to shorter periods. One of the observable consequences of
magnetic activity are star-spots, and their number, size, distribution
and variability, as deduced from surface images of CV secondaries,
would provide critical tests of stellar dynamo models in a hitherto
untested period regime.

However, determining whether features in Roche tomography
reconstructions are real or not is problematic as the resulting maps
are prone to both systematic errors, due to errors in the assumptions
underlying the technique, and statistical errors, due to measurement
errors on the observed data points. It is therefore essential to quantify
the effects of these errors on Roche tomograms in order to properly
assess the reality of any features present.
In this paper we study the artefacts produced by errors in the input
parameters using reconstructions of simulated datasets, and present a
technique for estimating the statistical errors on the reconstructions.
Subsequent papers will present the application of Roche tomography to
real datasets.

\section{Roche tomography}
\label{sec:imaging}

Roche tomography (\pcite{rutten94}; \pcite{rutten96}) is analogous
to the Doppler imaging technique used to map rapidly rotating single
stars (e.g. \pcite{vogt83}), detached binary stars \cite{ramseyer95}
and contact binaries \cite{maceroni94}. However, in contrast to the
Doppler imaging technique, the continuum is ignored in Roche tomography
because of the unknown and variable contribution of the accretion regions
to the spectrum, which forces Roche tomography to map absolute line
fluxes. Apart from this, the principles only differ in detail.

In Roche tomography we assume values for the binary parameters and that
the secondary is Roche-lobe filling, locked in synchronous rotation  and has a
circularised orbit. We then model the secondary as a series of
quadrilateral tiles of approximately equal area lying on the critical Roche
surface. Each tile or surface element is assigned a copy of the local
(intrinsic)
specific intensity profile convolved with the instrumental resolution.
These profiles are scaled to take into account the projected area,
limb darkening and obscuration and then Doppler shifted according to the 
radial velocity of the surface element at a particular phase.
Summing up the contributions from each element gives the rotationally 
broadened profile at that particular orbital phase.


In order to reconstruct a surface image of the secondary,
the `inverse' of this process must be carried out.
We do this by iteratively varying the strengths of the profile contributed
by each element assuming that the {\em shape} of the intrinsic line profile
does not change.
This iterative procedure is carried out (assuming no contribution from the
accreting regions) until a satisfactory fit to the
observed data, defined by the $\chi^2$ statistic (i.e. $\chi^2 \sim 1$),
is achieved. As there are a large number of different intensity distributions
that all predict trailed spectra consistent with the observed one, an
additional constraint is required. This is found by selecting the map of
maximum entropy with respect to a suitable default map and is performed
using the maximum entropy algorithm developed by \scite{skilling84}.
Further details on the Roche tomography process can be found in 
\scite{dhillon00}.

\section{systematic errors and artefacts}
\label{sec:simulated reconstructions}

\begin{figure*}
\begin{tabular}{ll}
\psfig{figure=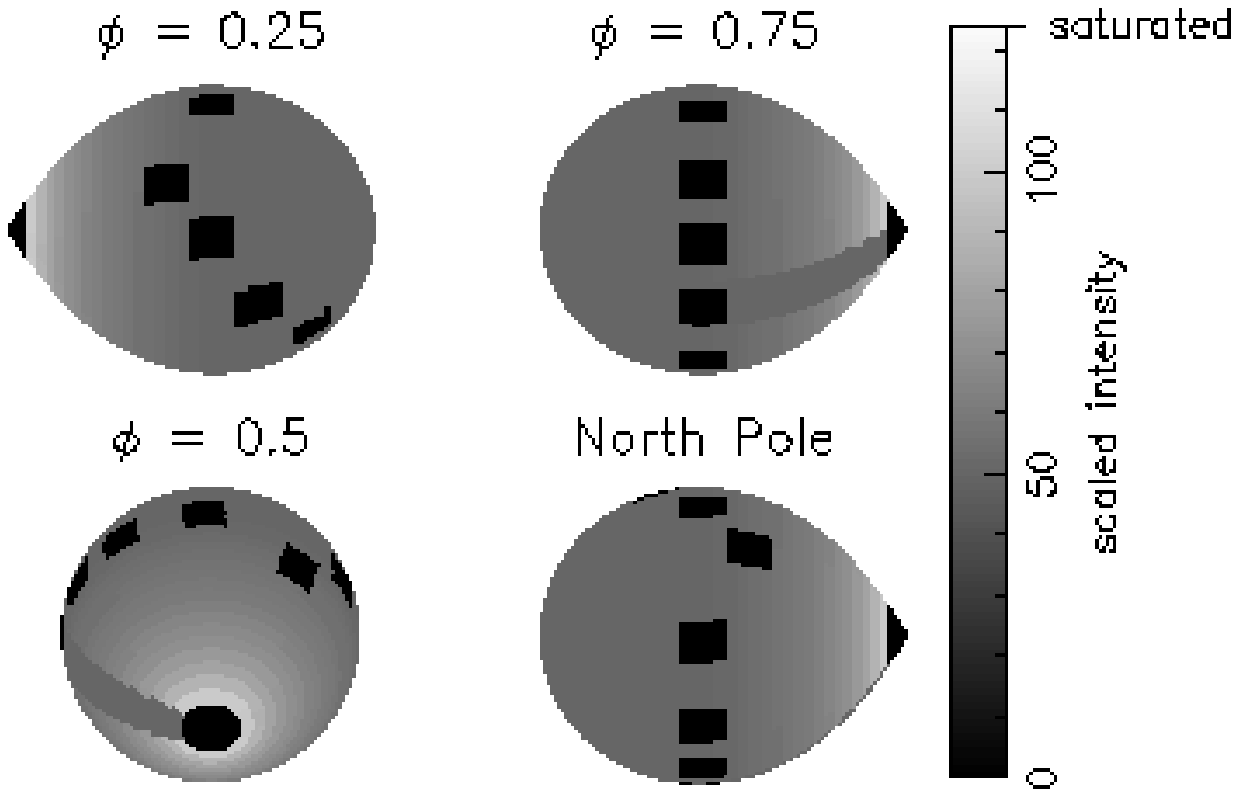,width=7.8cm} &
\psfig{figure=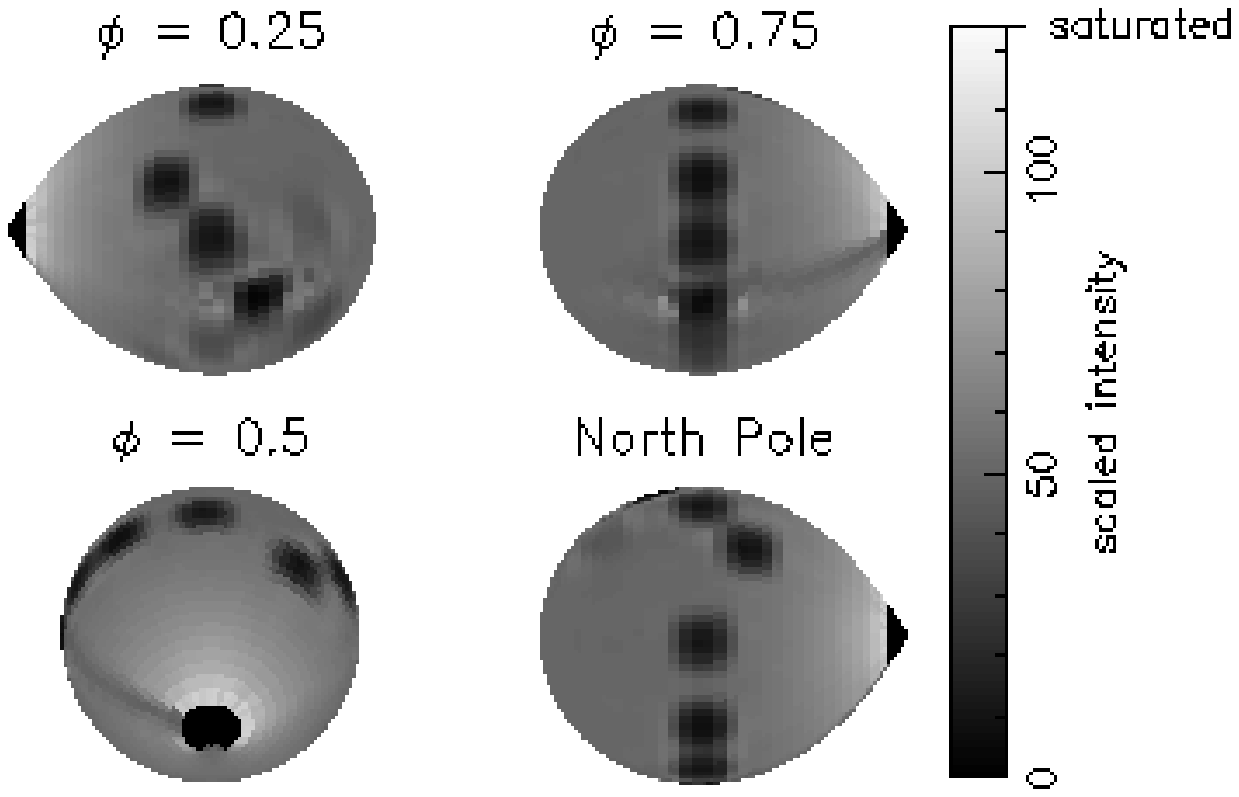,width=7.8cm} \\
Image A. The test image & Image B. Best fit
\vspace{0.45cm}\\
\psfig{figure=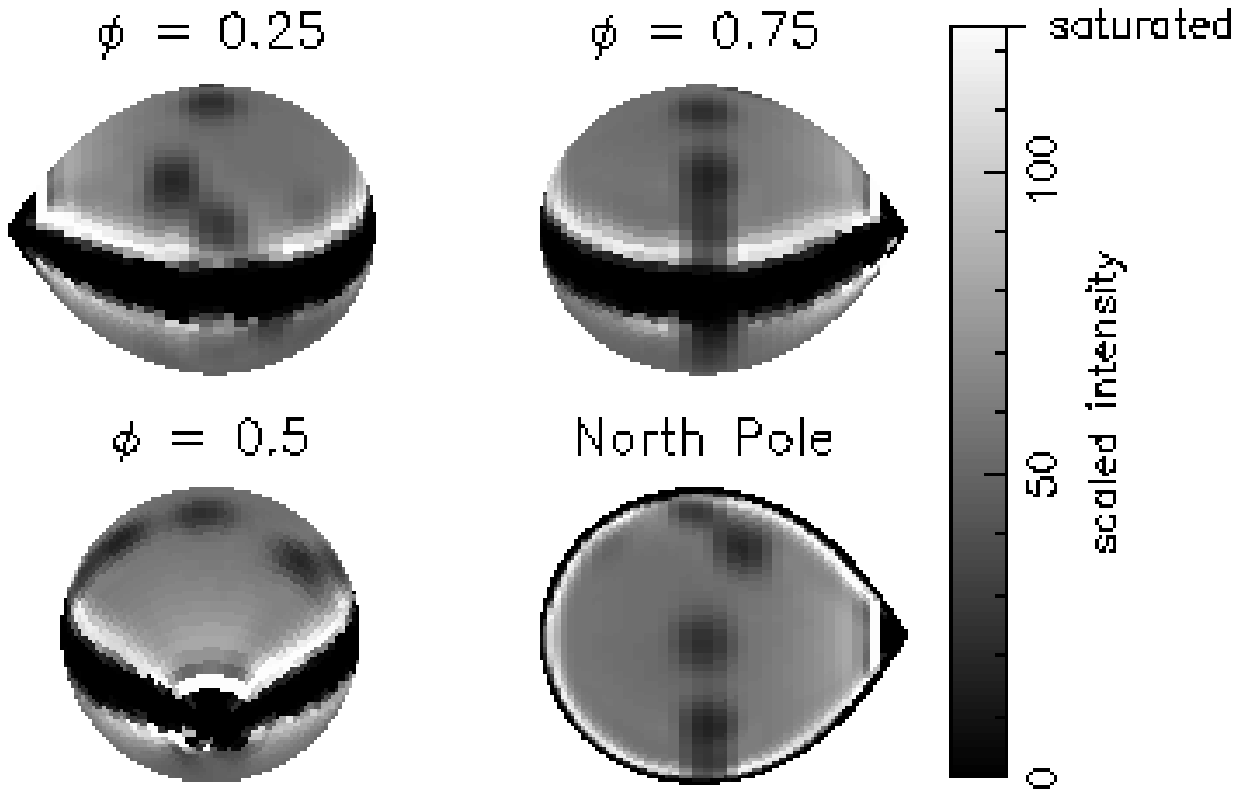,width=7.8cm} &
\psfig{figure=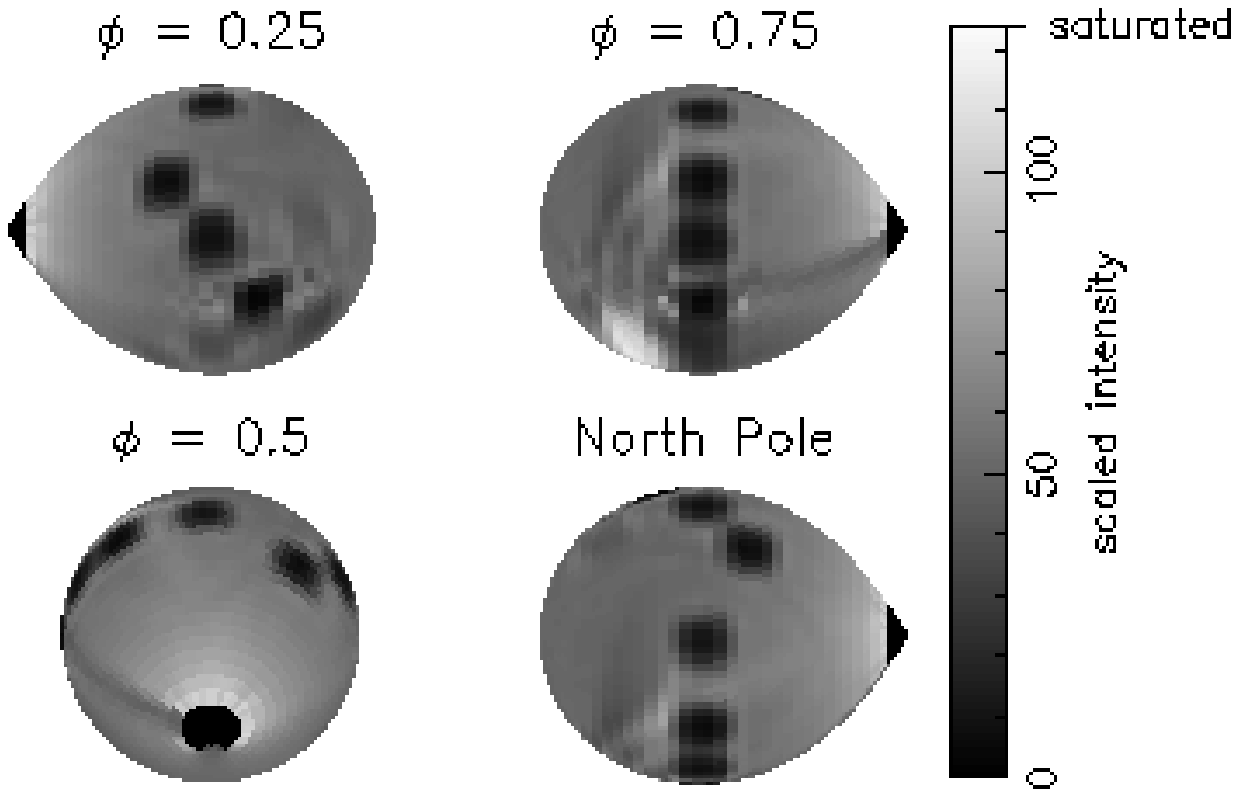,width=7.8cm} \\
Image C. Systemic velocity & Image D. Flare
\vspace{0.45cm}\\
\psfig{figure=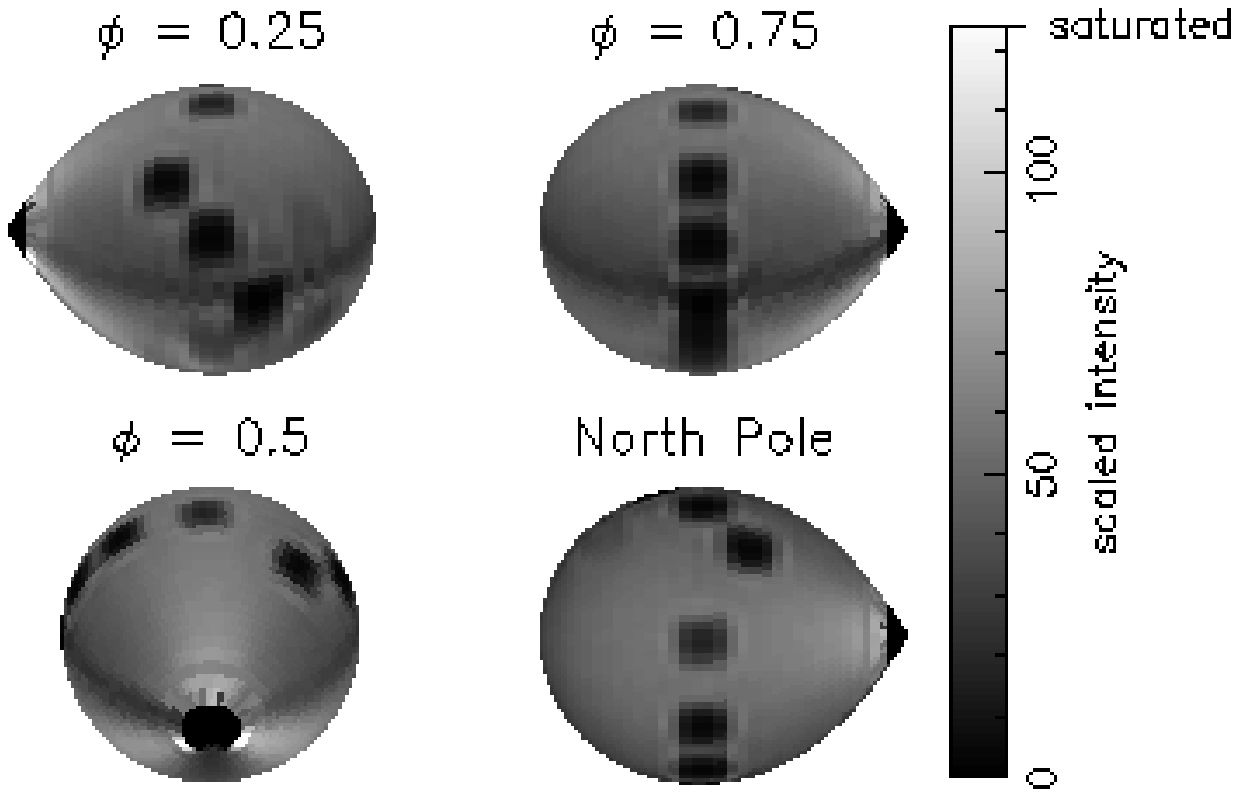,width=7.8cm} &
\psfig{figure=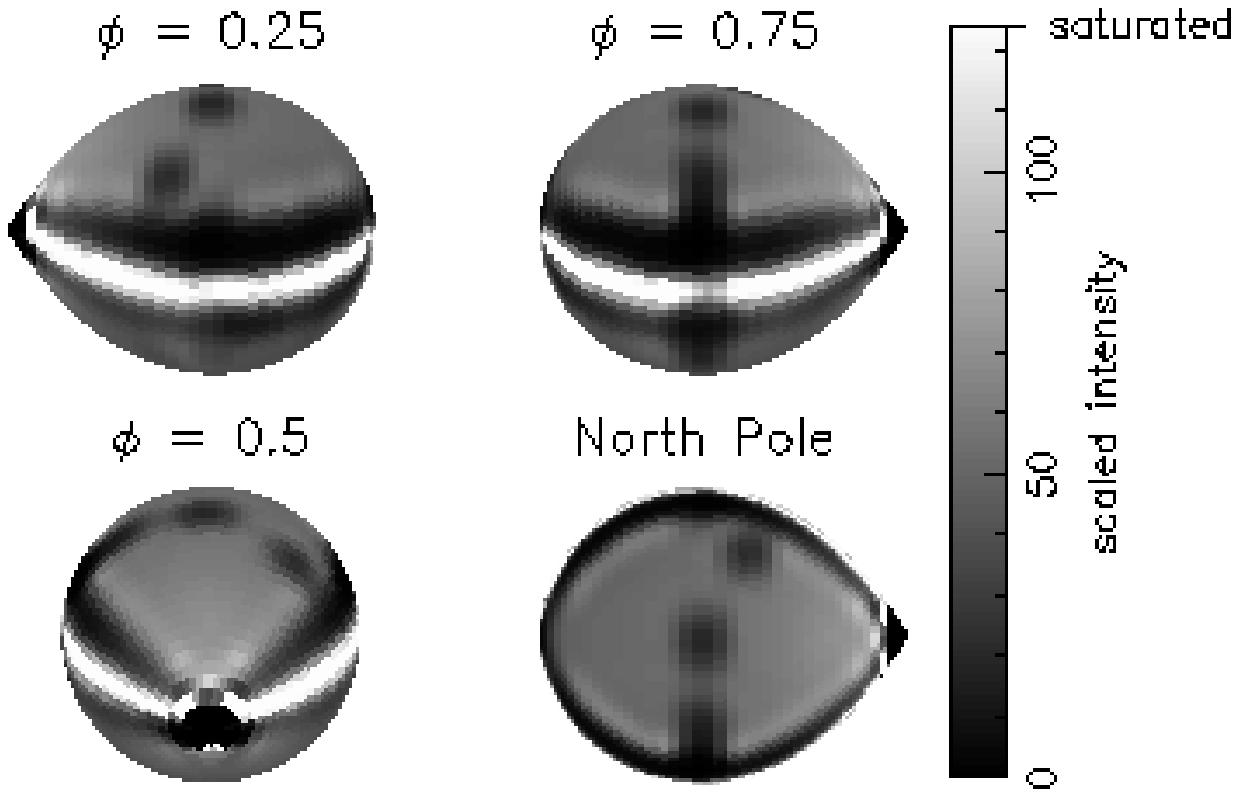,width=7.8cm} \\
Image E. Limb darkening & Image F. Velocity smearing
\vspace{0.45cm}\\
\psfig{figure=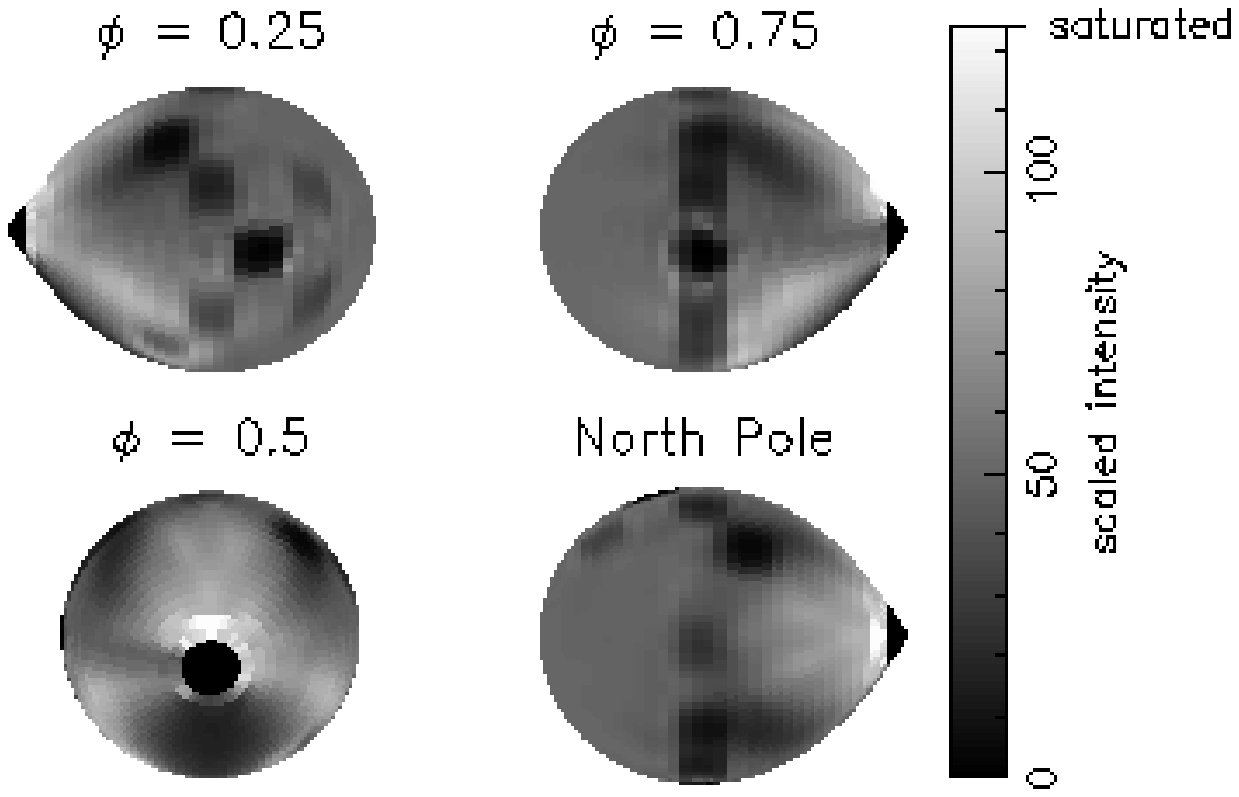,width=7.8cm} &
\psfig{figure=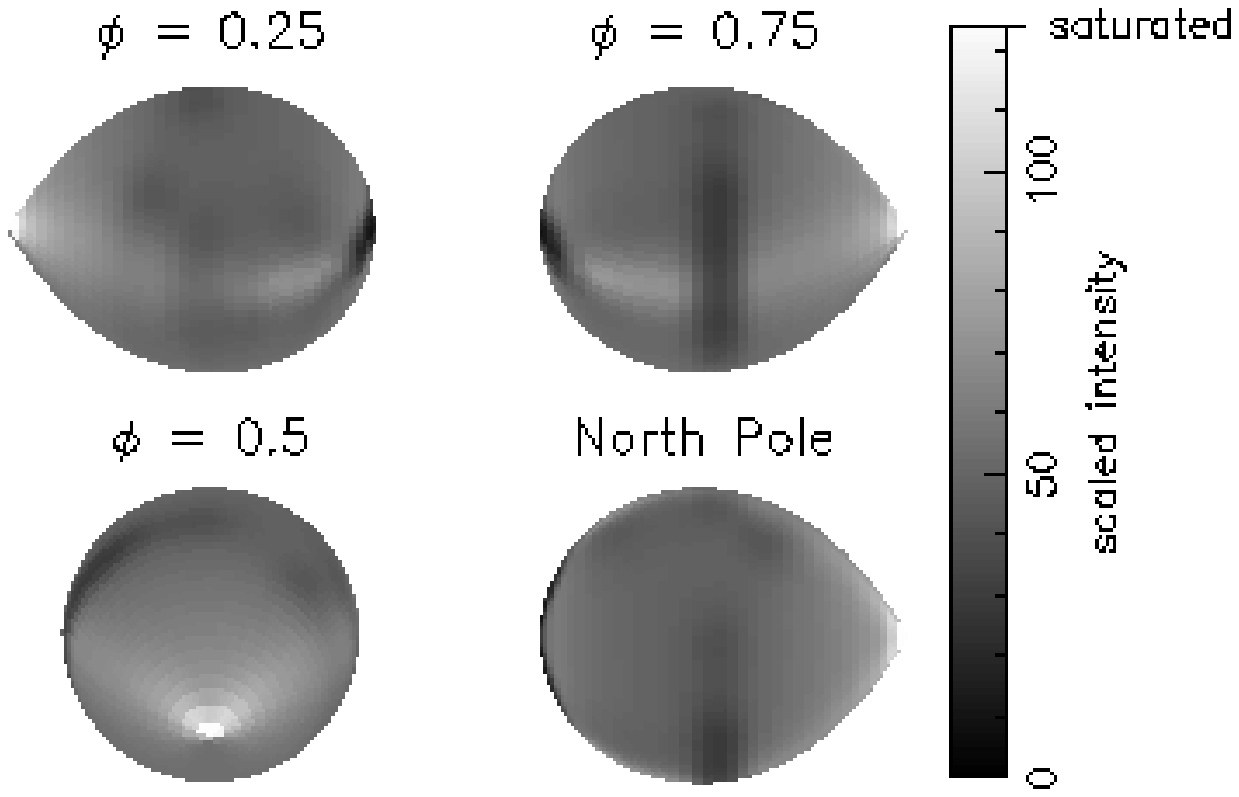,width=7.8cm} \\
Image G. Eclipse by an accretion disc & Image H. Incorrect secondary mass
\vspace{0.4cm}\\
\end{tabular}
\caption{The effects of systematic errors on the test image.}
\label{fig:systematics}
\end{figure*}

\begin{figure*}
\begin{tabular}{ll}
\psfig{figure=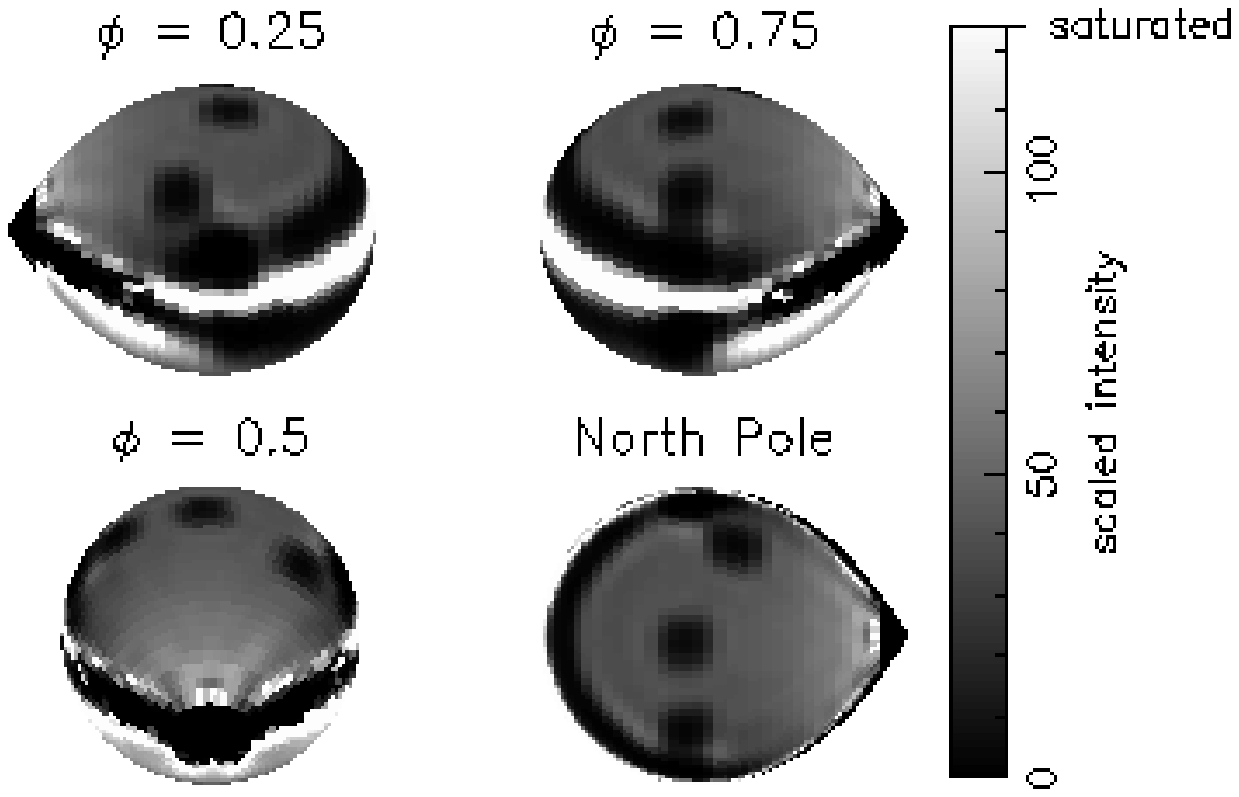,width=7.8cm} &
\psfig{figure=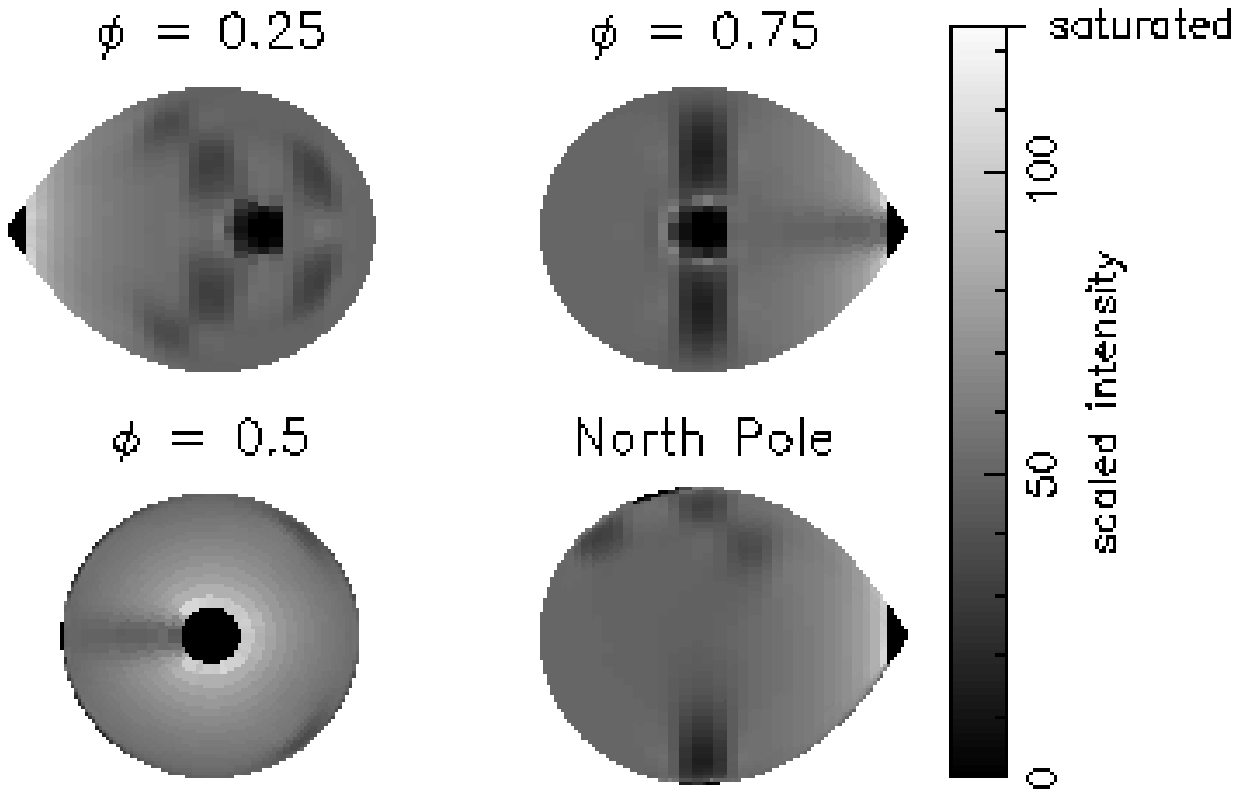,width=7.8cm} \\
Image I. Incorrect inclination & Image J. The `mirroring' effect
\vspace{0.45cm}\\
\psfig{figure=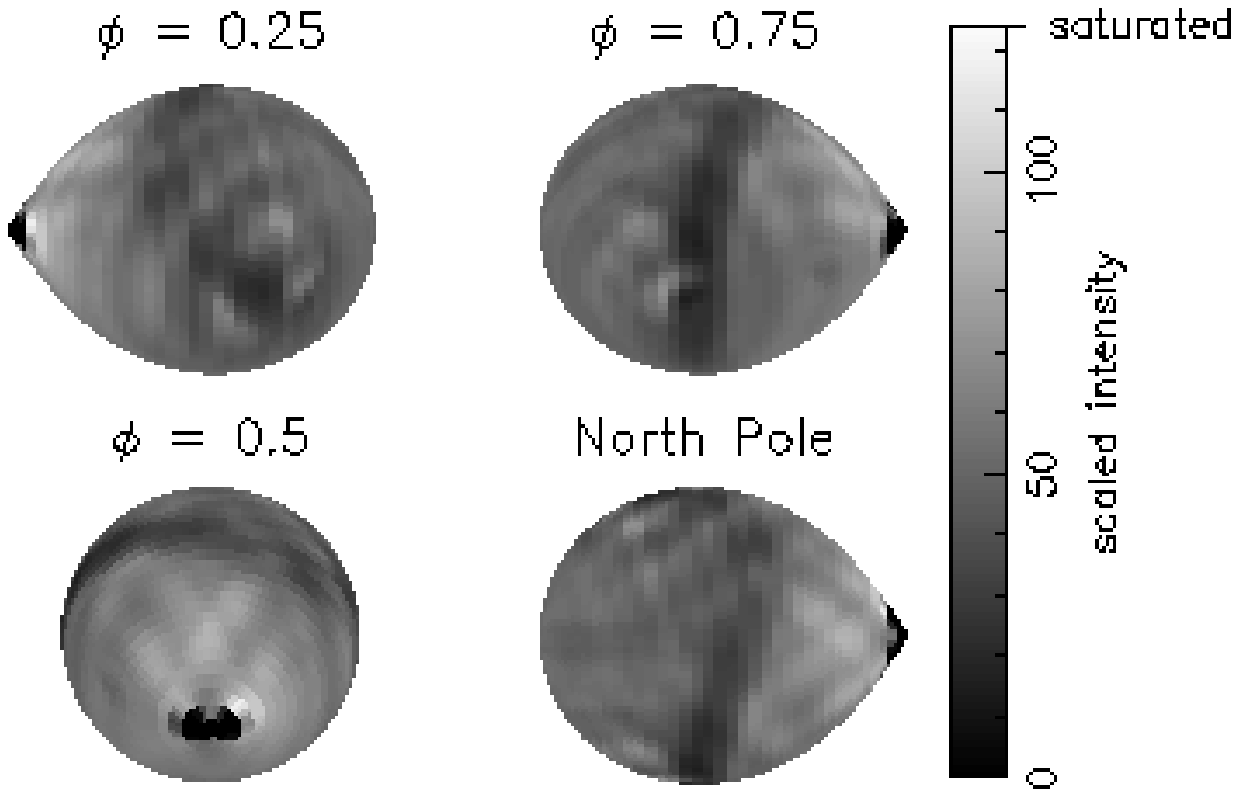,width=7.8cm} &
\psfig{figure=artefacts.ps,width=6.0cm,angle=-90} \\
Image K. Phase under-sampling & Image L. Ring-like streaks due to phase under-sampling
\vspace{0.45cm}\\
\psfig{figure=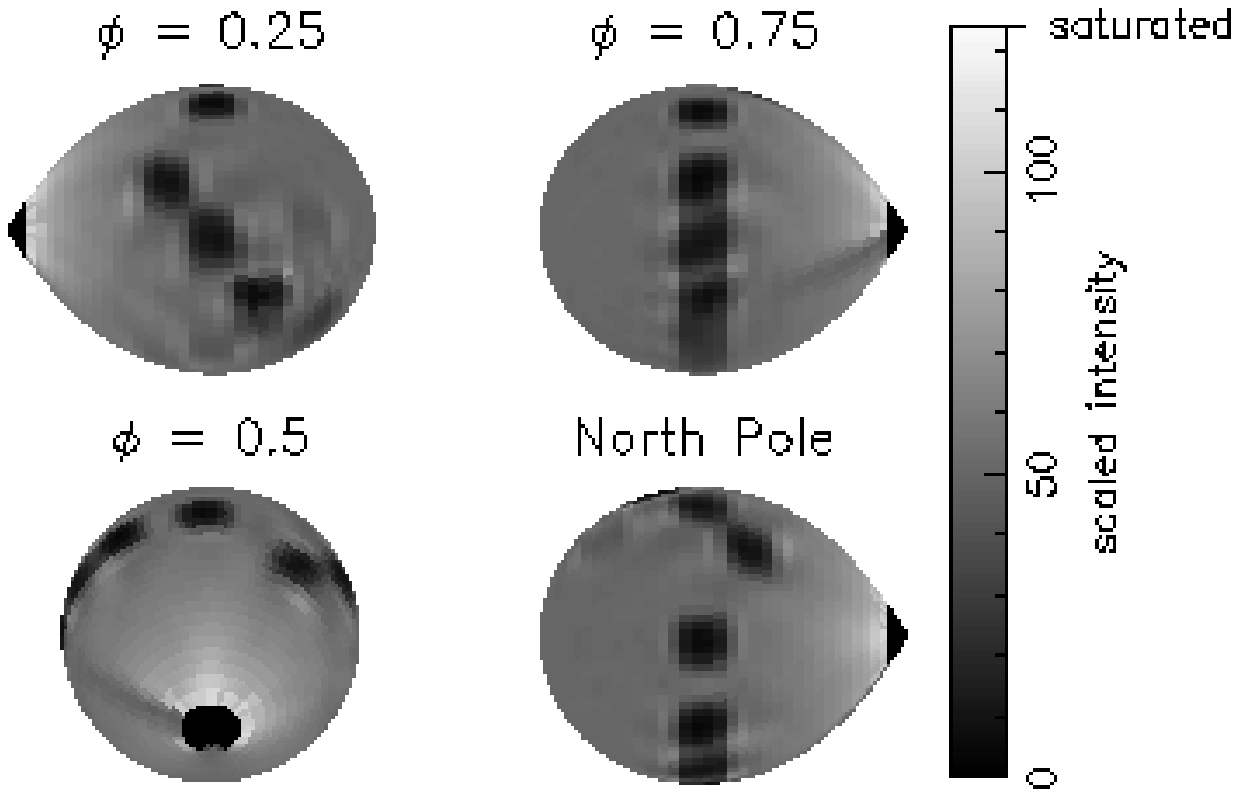,width=7.8cm} &
\psfig{figure=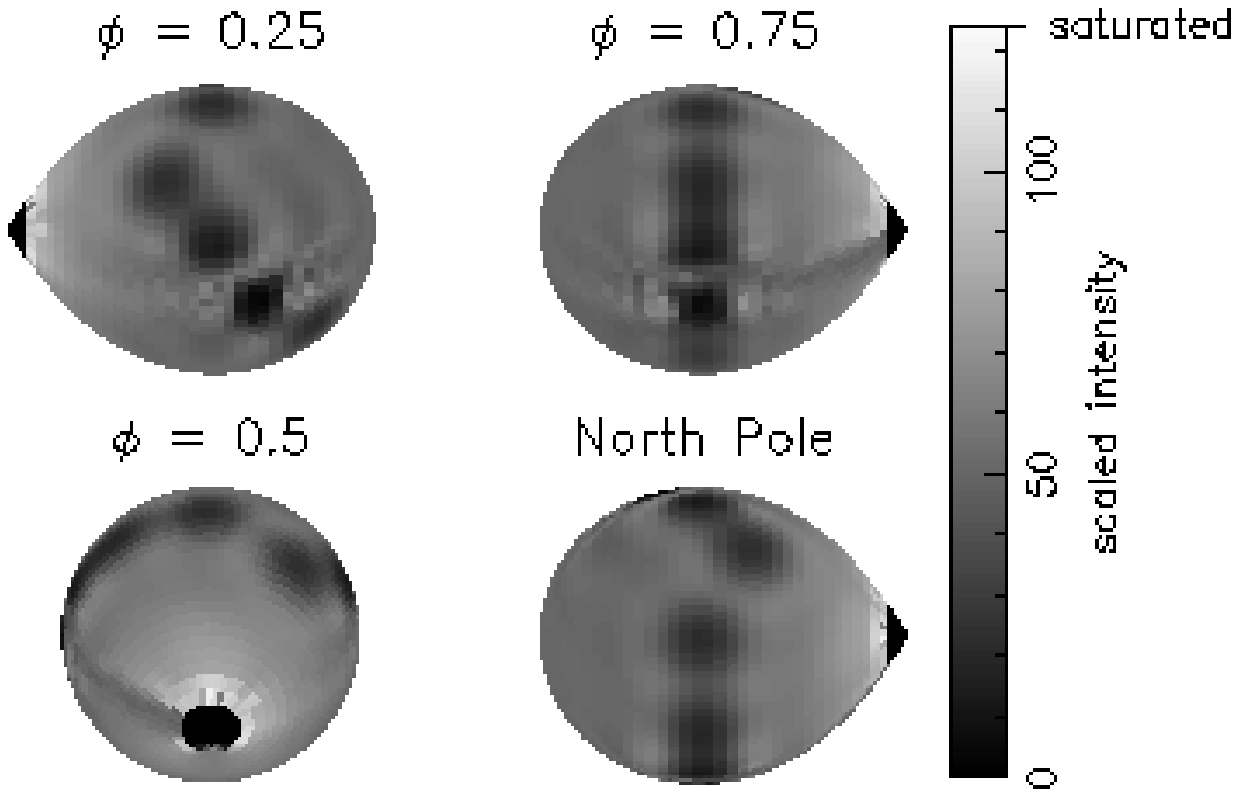,width=7.8cm} \\
Image M. Incomplete phase coverage & Image N. Resolution
\vspace{0.45cm}\\
\psfig{figure=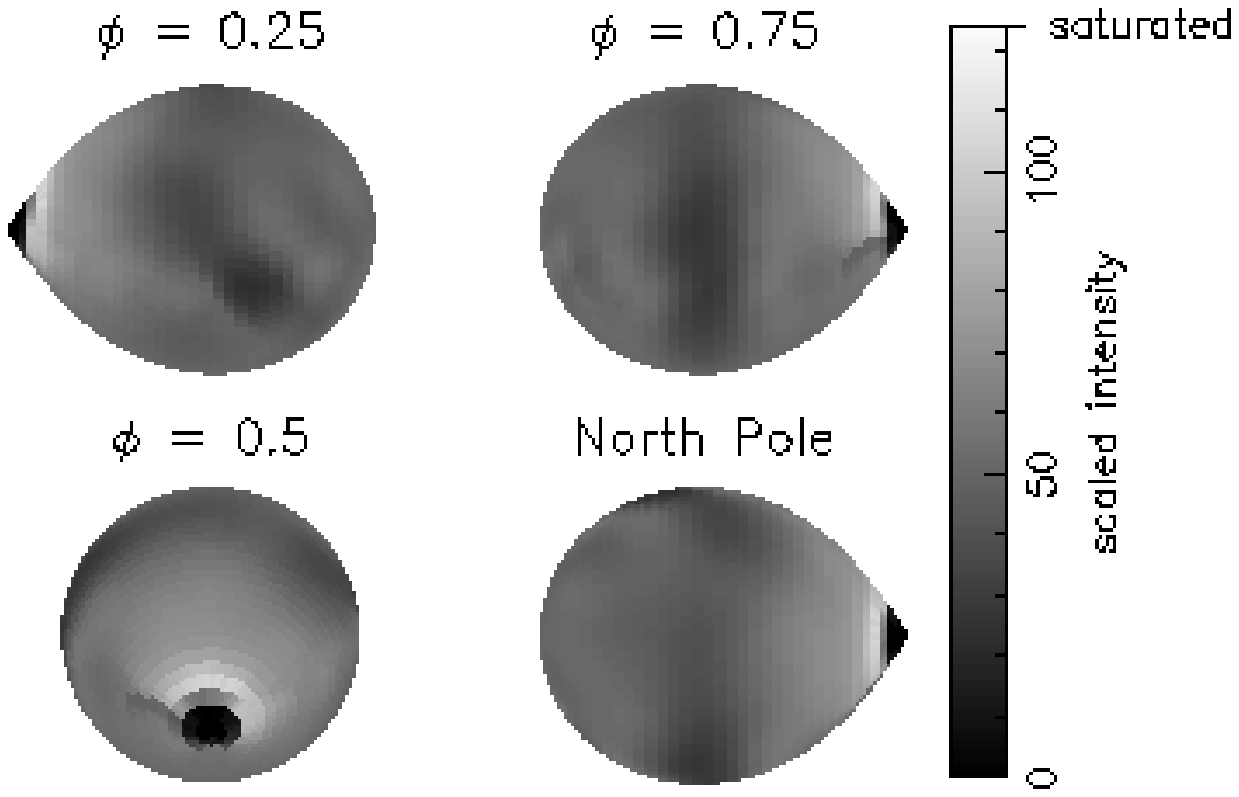,width=7.8cm} &
\psfig{figure=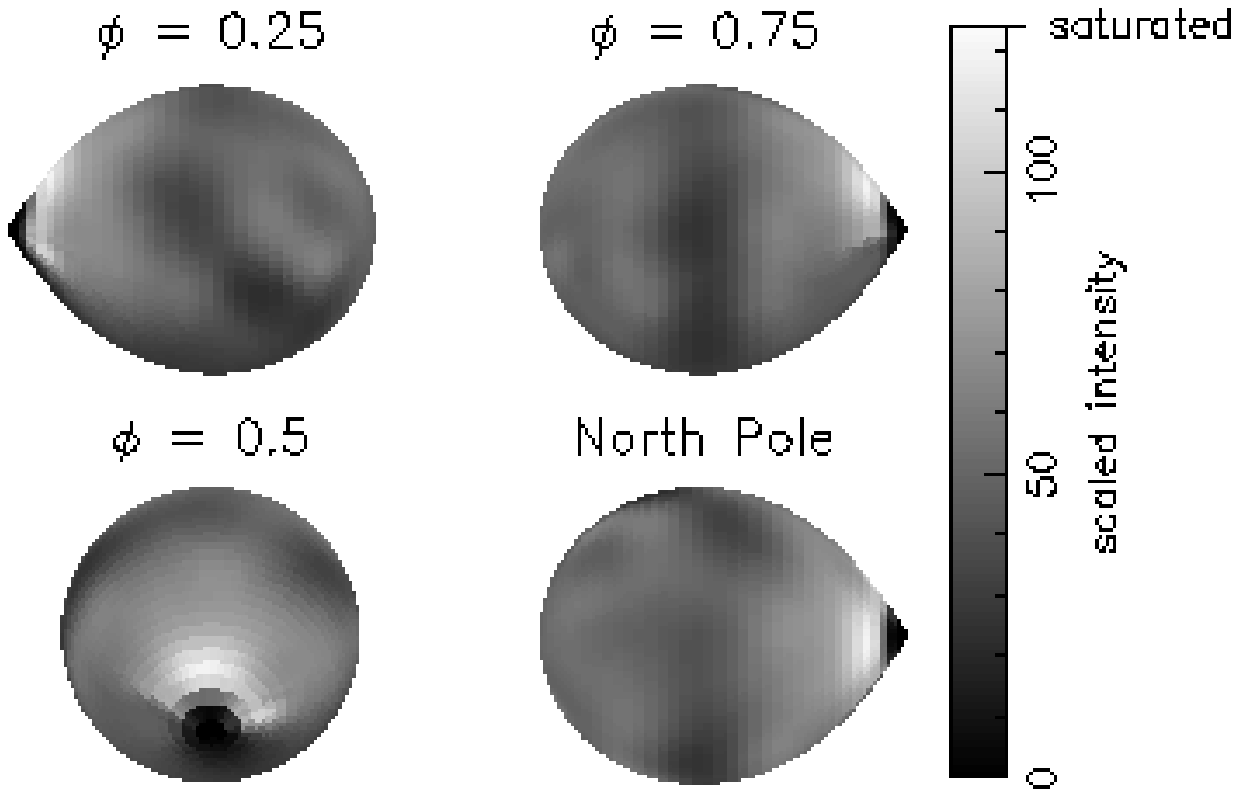,width=7.8cm} \\
Image O. Uniform default map & Image P. Gaussian-blurr default map
\end{tabular}
\vspace{0.4cm}\\
\centerline{{\bf Figure 1 (continued).} The effects of systematic errors on the test image.}
\end{figure*}

There have been a number of studies exploring the effects of 
systematic errors on Doppler images of single stars. For example, 
the significance of so-called polar spots (large, long-lived, high latitude 
spots straddling the polar
regions) has been tested for robustness against errors such as
incorrect line profile models \cite{unruh95}, chromospheric emission
(\pcite{unruh97}; \pcite{bruls98}), gravity darkening and differential
rotation \cite{hatzes96}. In this section we explore the effects of
systematic errors on Roche tomograms of CVs using simulated reconstructions
in a similar manner to \scite{marsh88b} in their treatment of systematic
errors on Doppler tomograms of CV accretion discs.

\subsection{The test image}

All our simulations are based on a single test image
(Image A, figure~\ref{fig:systematics}). This image mimics
the effect of irradiation of the secondary star by the compact primary,
taking into account the distance to the companion and the incidence
angle at the surface of the star. The intensity of the side of the star
that is not irradiated is set to the intensity at the terminator.
Also, shielding by (for example) an accretion
stream is mimicked by a narrow band stretching along the equator
from the inner Lagrangian (L$_1$) 
point to the terminator and covering 3.6\% of the
total surface area. The intensity of this band is equal to the intensity
of the non-irradiated side of the secondary.

In addition, an array of ten completely dark spots are distributed around
the star. Four are located on the trailing hemisphere centered at latitudes
of approximately +50$^\circ$, +25$^\circ$, 0$^\circ$ and --25$^\circ$,
all with different longitudes. Another four are located on the leading
hemisphere at latitudes of +50$^\circ$, +25$^\circ$, 0$^\circ$ and
--36$^\circ$, all with the same longitude. The two final spots are located
at the north pole (+90$^\circ$) and the L$_1$ point.
Each spot covers 0.7\% of the
total surface area of the secondary, with the exception of the L$_1$ spot
which covers 1\%.

Finally, the geometry of the Roche-lobe itself is given by the binary
parameters $M_1$ = 1.0 M$_\odot$, $M_2$ = 0.5 M$_\odot$ and an orbital
period of 2 hours. Unless otherwise stated the orbital inclination
used was 60$^\circ$. In total, the image consists of 3008 elements which,
with these binary parameters, gives a maximum radial velocity difference 
between any two neighbouring elements of 18.1 km\,s$^{-1}$.

\subsection{Synthetic datasets and fits}

\begin{figure}
\psfig{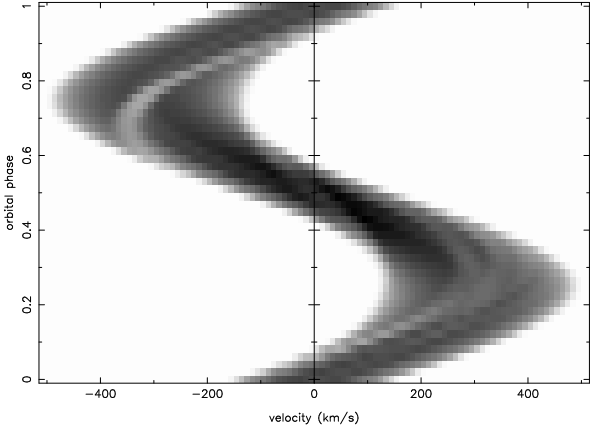}
\caption{Synthesized dataset produced by the test image.}
\label{fig:dataset}
\end{figure}

Unless otherwise stated, a perfect synthetic trailed spectrum was
computed assuming an instrumental resolution of 20 km\,s$^{-1}$ and
a negligible intrinsic line width. The whole orbit was sampled
over 50 evenly spaced phased bins, with a velocity bin-width of
10 km\,s$^{-1}$ and a negligible exposure time. No contribution from
an accretion disc or other part of the system was considered. A sample
trailed spectrum is shown in figure~\ref{fig:dataset} and the best fit
to this dataset is displayed in Image B, figure~\ref{fig:systematics}.

In each case the initial guess at the fit was one of uniform intensity
distribution and a uniform default map was used except in
section~\ref{sec:defaults}. Due to the nature of the tests, the fits
could not proceed to the same final $\chi^2$ on each occasion. Thus the
optimal fit was determined by minimising $\chi^2$ until the spot features
were resolved.

The reconstructions are plotted using a linear grey-scale. A value of 100
corresponds to the maximum intensity on the original test image (Image A,
figure~\ref{fig:systematics}).
Any intensity value $\sim22\%$ greater than the maximum intensity of the test
image is flagged as white. This is because some artefacts generate `hot'
elements with high intensities which would otherwise dominate the grey-scale
plots.

We are now ready to explore the effects of systematic errors on the Roche
tomography reconstructions. In each case a fake dataset was generated that
was either corrupted in some form or was fitted assuming an incorrect
parameter.

\subsection{Systemic velocity $\gamma$}
\label{sec:gamma}

The systemic velocity is the radial velocity of the centre-of-mass of
the binary system and represents the mean shift of the line from its
rest wavelength. In cases where the wrong value for the systemic
velocity is assumed, the reconstruction process will have difficulty
in fitting the line profiles. As a result, artefacts will be generated
in order to allow the model to fit the profiles as well as possible.
Due to the high quality of the synthetic datasets used,
one might expect a small error in the systemic velocity to produce
noticeable artefacts.
Image C (figure~\ref{fig:systematics}) shows the effects of ignoring a 
systemic velocity of +5 km\,s$^{-1}$
present in the data. Although the spot features and, to some extent, the
irradiated inner face of the secondary are still visible in the
reconstruction, dark bands framed by bright stripes are formed
around the equatorial regions.

These equatorial stripes can be explained as follows. When fitting the
trailed spectrum the data appears to be `skewed' to one side of the
radial velocity axis and therefore the model cannot possibly fit
the edges of the line profiles. At the lower (or most negative) radial
velocities expected by
the model there is no data, the data that should have been present at
these velocities having been shifted to a higher radial velocity. Thus
the model attempts to fit zero intensity to the regions of the star that
contribute to the line profile at these points, resulting in a dark 
equatorial band.

At the other extreme, the higher (or most positive) radial velocities in
the line profiles are shifted beyond the velocities expected by the model. 
Thus the intensities found at the highest radial velocities
are far greater than expected by the model and the opposite scenario
emerges. The model now attempts to fit high intensities to the regions
that contribute to the line profiles at these points. As only the
equatorial regions ever contribute to the edges of the line profiles, 
a bright equatorial band results. The algorithm does not allow negative
image values, which means it has greater freedom in setting the intensity 
(and hence the location) of the bright band. The bright band therefore
borders the dark band, which lies on the equator. 

\subsection{Time variability}
\label{sec:flare}

One of the key assumptions of our model is that features on the
surface of the secondary do not vary during the course of an 
observation, e.g. a spot does not suddenly appear/disappear at a
particular binary phase. Here we
explore how a variation in the line flux from one region of the secondary
may effect the reconstruction
by simulating the presence of a flare. The flare is
situated on the equator of the leading hemisphere (phase 0.75) and displaced
slightly towards the back of the star. It is only present between phases
0.8 and 0.92 inclusive, has approximately six times the intensity
of the surrounding non-irradiated region and covers 0.7\%
of the total surface area.

In the reconstruction (Image D, figure~\ref{fig:systematics}) 
a bright region can be seen at
phase 0.75. This region is at roughly the same longitude but at
a much lower latitude than the actual position of the flare. This shift
in latitude is easily explained; due to the orbital inclination, regions
near the south pole are visible for shorter periods than the equatorial
regions. By placing a bright region there, the fitting routine is trying
to account for the limited phase coverage over which the flare is
observed. However, in doing so the model must smear this feature
in order to match the observed radial velocities of the flare.
In addition, bright `ringing' can be seen, which is due to the
limited phase coverage over which the flare is observed; an explanation
of this effect will be given in more detail in section~\ref{sec:phase}.

\subsection{Limb darkening}
\label{sec:limb}

In section~\ref{sec:imaging} we mentioned that the line profile is scaled
to take into account limb darkening. The fitting procedure cannot, without
prior information, account for limb darkening. This is because limb
darkening causes each element in the map to have a phase-dependent intensity 
variation. We must therefore supply the fitting procedure with an appropriate
limb darkening relation. In this section we show
the effects of assigning incorrect limb darkening strengths and laws.

First, data was generated assuming a linear limb darkening coefficient
of 0.5. This was then fitted assuming no limb darkening (Image E,
figure~\ref{fig:systematics}). The spot
features are readily visible, although a dark equatorial band is formed.
This band can be understood because the model calculates much higher 
intensities at the edges of the profiles than are present in the fake 
dataset. As described in section~\ref{sec:gamma}, this results in
dark equatorial bands in the reconstruction. In addition, the overall 
intensity level of the map is less than that of the test image due to the 
reduction in light caused by the limb darkening. Conversely, if one assumes 
too high a limb darkening coefficient then the opposite occurs.
The model calculates much lower intensities at the edges of the profiles and 
in order to compensate the fitting routine produces bright bands around the 
equator.

For our second test, we assumed an incorrect limb darkening law. Data
generated using a linear limb darkening coefficient of 0.5 was then fitted
assuming quadratic limb darkening co-efficients of $a$ = --0.025,
$b$ = 0.635 and

\[I\left(\mu\right) = I_1 \left(1-a\left(1-\mu\right) - b\left(1-\mu\right)^2\right),\]
where $\mu$ = $\cos\gamma$, $\gamma$ is the angle between the line of
sight and the emergent flux and $I_1$ is the monochromatic specific intensity at $\mu$=1. The reconstruction showed a bright band 
around the equatorial regions where the fitting routine has had to compensate 
for the increased limb darkening predicted by the quadratic limb 
darkening law in this case.

\subsection{Velocity Smearing}
\label{sec:smear}

Velocity smearing is the degradation in spectral resolution due to
the combined effects of a finite exposure time and the orbital motion of 
the secondary star. Velocity smearing can be ignored if it is 
insignificant compared to the instrumental resolution. To obtain
adequate signal-to-noise, however, exposure times, $t_{\rm exp}$, are 
usually selected which match the resulting velocity smearing to the 
instrumental resolution, $v_{\rm res}$, via the formula

\begin{equation}
t_{\rm exp} \sim P v_{\rm res} / 2 \pi K_R,
\label{eqn:smearing}
\end{equation}
where $P$ is the orbital period of the binary and $K_R$ is the radial-velocity
semi-amplitude of the secondary star. In cases where longer exposure
times are unavoidable, significant velocity smearing will occur.
It is possible to correct for the effects of velocity smearing 
in Roche tomography by calculating the line profiles at intermediate 
phases during an exposure and averaging the results. This slows
the iterations, however, so in this section we explore what happens if 
velocity smearing is neglected during the reconstruction process.

\begin{figure}
\psfig{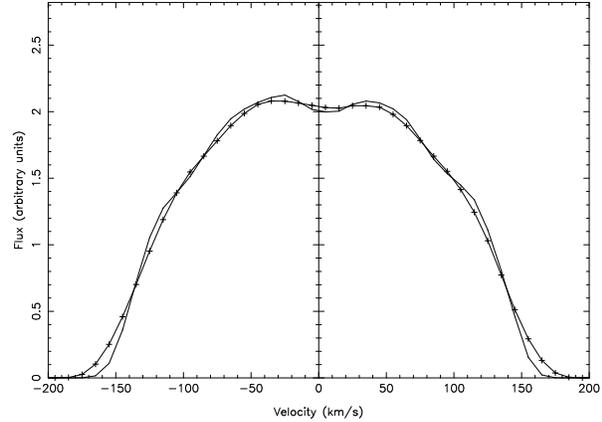}
\caption{Solid line: line profile at phase 0, computed assuming
zero exposure time. Crossed line: line profile around phase 0, computed
assuming an exposure time twice that required by equation~\ref{eqn:smearing}.}
\label{fig:spectrum}
\end{figure}

Image F (figure~\ref{fig:systematics}) shows what happens when an exposure 
time twice as long as that required by equation~\ref{eqn:smearing} is assumed
when generating fake data and then ignored in the reconstruction process.
Once again the largest effects are seen in the equatorial regions.
Figure~\ref{fig:spectrum} shows why this occurs. The radial
velocities of the edges of the smeared profile extend beyond the 
velocities expected when the exposure time is ignored. For the same
reasons as discussed in section~\ref{sec:gamma}, 
this results in bright and dark artefacts around the equator.
Figure~\ref{fig:spectrum} also shows how bumps in the line profile
due to star-spots (e.g. the dip around 0 km\,s$^{-1}$) are smeared,
resulting in a blurring of the star-spots in the reconstruction.

\subsection{Intrinsic line profile width}
\label{sec:profile}

We have so far assumed that the intrinsic line-profile width
is negligible in comparison with the rotational broadening. This
may not always be the case and the width or shape of the intrinsic 
line profile may be important. Here we examine the effects of assuming 
an incorrect intrinsic line-profile width.

Data was produced assuming 20 km\,s$^{-1}$ instrumental resolution as
before but with an intrinsic profile width of 45 km\,s$^{-1}$ (FWHM). The
reconstruction was carried out assuming no intrinsic line-profile
width and the result was very similar to Image F
(figure~\ref{fig:systematics}). Once more the edges of the lines
cannot be fitted and banding is hence seen around the equator. The
reconstruction also shows that the spots are broadened. This is due to
the algorithm attempting to match the width of the bumps in the line
profile which, in the absence of intrinsic broadening, it can only do
by increasing the size of the spots in the reconstruction.

\subsection{Eclipse by an accretion disc}
\label{sec:disc}

The highest inclination CVs may show signs of an eclipse of the
secondary star by an accretion disc around phase 0.5. This can
be accounted for during the reconstruction process either by removing
the affected spectra (as commonly done in Doppler tomography during
primary eclipse), or by including the eclipse in the algorithm.
If neither of these steps are taken, artefacts will be introduced in
the reconstructions, as described in this section. 

A model including a cylindrical accretion disc of radius $0.5 R_{L_1}$ and
height $0.1 R_{disc}$ was created and a synthetic trailed spectrum including
the eclipse by the disc was generated assuming $i=80^{\circ}$.
Image G (figure~\ref{fig:systematics}) shows the reconstruction if the 
eclipse by the accretion disc is ignored. As one might expect, 
the artefacts generated by the
eclipse are most apparent around the L$_1$ point. In particular,
the southern regions (where light is blocked by the eclipse) appear dark,
and there are alternating regions of bright and dark bands radiating from
the L$_1$ point. These are due to the attempts of the fitting algorithm 
to match the eclipse phases, resulting in a reduction of intensity on
the inner hemisphere which is compensated for at other phases by the bright 
bands. Note that the spot features are well reconstructed, but exhibit the
mirroring effect described in section~\ref{sec:mirror} due to the high
inclination.

\subsection{Masses}
\label{sec:masses}

A knowledge of the binary masses are essential in order to 
define the geometry of the secondary star model. Very few accurate
mass determinations are available for CVs, however, and though we can use 
Roche tomography to determine them (see section~\ref{sec:eland}) we still 
need to know how an error in the masses may effect the reconstructions.

Image H (figure~\ref{fig:systematics}) shows the effect of underestimating 
the secondary mass by
0.1M$_{\odot}$. The effect of reducing the secondary's mass whilst keeping
the period and inclination the same is to shift the model to higher
radial velocities. Hence the spot features on the reconstruction
appear shifted towards the L$_1$ point and lower radial velocities in order 
to compensate for this. In addition, as the radial velocities of the spot
features in the original dataset no longer correspond to a well defined
location on the secondary, the spots appear smeared over a larger area.
At phases 0.25 and 0.75 the outer hemisphere of the model is shifted to
higher radial velocities than present in the dataset and hence a dark band
is formed near the rear of the star. At other phases the model cannot
match the width of the line profiles due to the reduced radius of the star
and this results in a bright equatorial band.

\subsection{Inclination}
\label{sec:incl}

Up until now we have always assumed the correct orbital
inclination (60$^{\circ}$) in each of our tests. In
reality, however, we may not know the inclination accurately.
Image I (figure~\ref{fig:systematics})
shows the effect of assuming an incorrect inclination of 55$^{\circ}$.

Once again, the most obvious artefacts appear at the equatorial regions. The
lower orbital inclination shifts the model to lower radial velocities.
As a consequence, the greatest mismatch between the
observed and computed trailed spectra occur around phases 0.25 and 0.75.
At the lowest radial velocities expected by the model there is no data and
a similar scenario occurs to that discussed in
section~\ref{sec:gamma}; dark bands are fitted around the equator
on the inner hemisphere of the star to compensate for the lack of data
at these velocities. Bright regions then appear around the dark
band in order to fit the intensities seen at other phases.
At higher radial velocities around phases 0.25 and 0.75,
the observed intensity is far greater than expected by the model, and the
opposite occurs on the outer hemisphere of the star.

The spot features remain visible, but are systematically shifted
towards the outer hemisphere of the star. This shifts them to regions of the 
star with higher radial velocities, which is necessary to counteract the 
shift of the model to lower radial velocities caused by assuming an incorrect
inclination. The opposite occurs if too high a value for the 
orbital inclination is assumed; the dark bands become bright and 
the spot features are systematically shifted towards the inner
hemisphere and lower radial velocities.

\subsection{The mirroring effect}
\label{sec:mirror}

The following sections describe artefacts that are not strictly due to
systematic errors but are a result of factors that are either
within our control (e.g. instrumental resolution, phase sampling) or are
inherent to the reconstruction process itself.

We begin with the problem that, although the radial velocities
contain latitudinal information, they cannot constrain whether a feature
is located in the northern or southern hemisphere. This information can
only be supplied by self-obscuration, but this becomes ambiguous at higher
inclinations until, at an inclination of 90$^{\circ}$, a feature 
in the northern hemisphere is self-obscured in an
identical manner to a feature located at the same latitude in the
southern hemisphere. This results in a `mirroring' of features in both
hemispheres as demonstrated in Image J (figure~\ref{fig:systematics}),
which also shows that the features become less apparent due to the fact
that they are smeared over twice the area they originally covered.

\subsection{Phase sampling}
\label{sec:phase}

Up until now our simulations have been carried out over 50 evenly
sampled phase bins covering a whole binary orbit. In practice, however,
it may not always be possible to obtain this number of phases due to
signal-to-noise requirements, or there may be gaps in the orbital coverage
because of weather conditions. Here we model such scenarios.

First we explore how under-sampling in phase can affect the reconstruction.
Image K (figure~\ref{fig:systematics}) was reconstructed from 6 evenly spaced
phases (5 independent) using zero exposure lengths\footnote{Although such a
scenario would probably only occur if long exposure times were being used, we
have  assumed zero exposure length in order to separate the
effects of velocity smearing (described in section~\ref{sec:smear}) and
those due to phase under-sampling.} and it is dominated by 
ring-like streaks. The effect is analogous to the streaks observed in
Doppler tomography~\cite{marsh88b} and can be understood by
considering that lines of constant radial velocity on the secondary
star at a particular phase can be integrated along to construct a line
profile.  These lines of constant radial velocity are ring-like in
shape and if there are only a few phases, or if the profile is
particularly bright at a certain phase (as in the case of a flare;
section~\ref{sec:flare}), the streaks will not
destructively interfere, leaving ring-like artefacts on the Roche
tomogram. To demonstrate this effect 
we produced maps where those elements with the same
radial velocity as the spot features were represented as a grey-scale whilst
all other elements were left blank. 
This was done for each of the 6 phases in the
dataset and the resulting images were then superimposed on top of one another
so that regions where streaks overlap are shown darker 
(Image L, figure~\ref{fig:systematics}). It can be seen that the features in 
this image closely resemble the artefacts in Image K
(figure~\ref{fig:systematics}).

Image M (figure~\ref{fig:systematics})
shows the effect of incomplete phase coverage, with phases between
0.4 and 0.6 missing (which may occur if certain phases are discarded due
to, for example, an eclipse by the disc; see section~\ref{sec:disc}).
There is no problem recovering the features around
the L$_1$ point where the majority of the information is missing.
Although the other spot features are also readily recovered they
appear elongated or streaked due to the same process described in the above
paragraph.

\begin{figure}
\label{fig:artefact}
\end{figure}

\subsection{Instrumental resolution}
\label{sec:resolution}

The spectral resolution of the data governs the size of the features that 
can be imaged, as shown by the blurred spots in Image N 
(figure~\ref{fig:systematics}) where we have reconstructed the test image 
using data of lower resolution (60 km\,s$^{-1}$) than before.
The number of surface elements in the model should be chosen to
match the spectral resolution. If the model has too few elements, 
the separation between adjacent elements in velocity space (particularly 
around phases 0.25 and 0.75) is greater than the resolution of the data. 
This means that there will be parts of the line profile to which no surface 
elements in the model contribute (at a particular phase) and no acceptable
fit will be found. If, on the other-hand, too many elements in the model are 
used, there will be no problem in fitting the data but it will slow the 
reconstruction process.

\subsection{Default maps}
\label{sec:defaults}

In the reconstruction process the final map is the one of maximum entropy
(relative to an assumed default map) which is consistent
with the data. If the data are good then the final map is constrained by the
data and will not be strongly influenced by the default. In this case
the choice of default makes little difference to the final map. If the data
are noisy, however, the data constraints will be weak and the map will be
strongly influenced by the default. The default can therefore be
regarded as containing prior information about the map, and regions of
the map that are not constrained by the data will converge to the default
value. 

To investigate the effect of different default maps on the reconstruction, 
it is necessary to degrade the quality of the test dataset. We therefore
generated data over 15 phase bins with a resolution of only 
60 km\,s$^{-1}$  and a S/N ratio of only 50. The dataset was then fitted 
using a uniform default map and a Gaussian-blurr default map. The
former was created by setting every element in the default to the average 
value of the map, and hence constrains large-scale surface structure. The 
latter was created by smoothing the map using a Gaussian-blurring function 
and hence constrains small-scale surface structure. 

Image O (figure~\ref{fig:systematics}) shows the reconstruction using
the uniform default map and, as expected, it can be seen that the spots are 
more blurred than in the reconstruction using the Gaussian-blurr default map
(Image P, figure~\ref{fig:systematics}). Features below --60$^{\circ}$ 
latitude are never visible and hence in the fitting routine they are assigned
a value determined by the default map. In the
case of the uniform default map, this is the average element value of
the map. In the case of the Gaussian-blurr default map, however, 
the value assigned depends upon the intensity values of the neighbouring 
elements and, for elements that are not visible, this is close to zero 
intensity. This explains why the 
reconstruction using the Gaussian-blurr default 
has a dark southern limb and, to compensate for this, a bright region above
the L$_1$ point. Hence, although the use of a Gaussian-blurr default is more 
likely to resolve small features, it is important to be aware of the effects 
that visibility may have on the final outcome.

\section{Determining physical parameters: The entropy landscape}
\label{sec:eland}

If the centre-of-light and centre-of-mass of the secondary are not
coincident,  the star's radial-velocity curve will be distorted in
some way from the pure sine wave which represents the motion of its
centre-of-mass. The observed radial-velocity curves of CV secondaries
suffer from this distortion, due to both geometrical effects caused by
the Roche-lobe shape and non-uniformities in the surface distribution
of the line strength due to, for example, irradiation. If these factors
are ignored, they will lead to systematic errors in the determination
of the binary star masses.

By modelling the shape and mapping the surface intensity distribution of
the secondary star, Roche tomography can provide constraints on the CV
masses. This is because, as discussed in 
section~\ref{sec:masses}, incorrect values of $M_1$ and $M_2$ will
introduce artefacts in the reconstructions which will lower the entropy
of the final solution. The correct values of the component masses are 
therefore those that produce the map of highest entropy.

\begin{figure}
\psfig{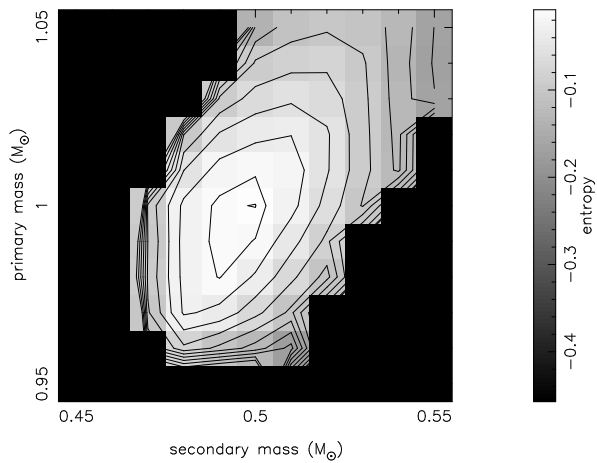}
\caption{Entropy landscape of the test data presented in 
figure~\ref{fig:dataset}. The contours are lines of constant entropy
and have been plotted to highlight the ridge of high entropy.}
\label{fig:entropyland}
\end{figure}

This is demonstrated in figure~\ref{fig:entropyland}, which displays
the `entropy landscape' of the test data shown in figure~\ref{fig:dataset}.
Each square in the entropy landscape corresponds to the maximum entropy
obtained in a reconstruction assuming a particular pair of component mass
values, a uniform default map and iterating to the same $\chi^2$ value
on each occasion. The reconstruction with the highest entropy was obtained 
for component masses of $M_1=1.0$ M$_{\odot}$ and $M_2=0.5$ M$_{\odot}$,
which are identical to the masses used to construct the test data. 
This shows the entropy-landscape method to be an extremely effective method
of determining accurate component masses which accounts for both geometrical
distortion and complicated intensity distributions on the secondary star. 
Note that the correct inclination was used for every reconstruction in
figure~\ref{fig:entropyland}, but in practice the inclination may not be 
accurately known. A
similar method to the entropy landscape can then be used to determine the
inclination or, in cases where the secondary eclipses the white dwarf,
the inclination can be varied for each combination of component masses in
order to remain consistent with the observed eclipse width.

Figure~\ref{fig:entropyland} also shows a ridge of high entropy
running along a line of (almost) constant $q$. Component masses lying
along this diagonal are therefore more difficult to distinguish between
than component masses lying perpendicular to it. The gradient of the diagonal 
can be derived by considering that the best fit is obtained when the 
centre of the model line profile matches the centre of the observed profile. 
The centre of the line profile is governed by the radial velocity of the
secondary, which relates directly to the distance of the
centre-of-mass of the secondary from the centre-of-mass of the binary, $a_2$. 
Thus only those combinations of $M_1$ and $M_2$ that maintain constant $a_2$
can provide satisfactory fits to the data. Using Kepler's law and assuming
a constant period and $a_2$ we can show that the ridges in the entropy
landscape correspond to component masses which obey the relation
\begin{equation}
\frac{M_1}{\left( M_1+M_2 \right)^{2/3}} = {\rm constant}.
\label{eqn:gradient}
\end{equation}
This simple argument is complicated in the case of irradiation.
\scite{rutten94} found that the ridge of high
entropy was shifted towards slightly higher values of $M_2$. Their
explanation for this was that the intensity distribution was strongly
peaked towards the L$_1$ point, causing a large difference with the uniform
default map. This difference was reduced by increasing $M_2$, allowing the
bright region to shift away from the L$_1$ point and to spread over more
elements, reducing the deviation between the map and the default whilst still
fitting the trailed spectrum to within its uncertainties. Despite this, the
mass determination was still correct to within $\sim$2 per cent.

Another feature of note in the entropy landscape of 
figure~\ref{fig:entropyland} is that, as the primary mass 
increases, there is a larger range of secondary masses that can be fitted 
satisfactorily. This can be explained by generalising 
equation~\ref{eqn:gradient} to
\[q + 1 \propto M_1^{1/2},\]
which shows that as the primary mass increases, the mass ratio, $q=M_2/M_1$, 
must also increase in order to hold $a_2$ constant. It can be shown that the 
secondary star radius is related to $q$ and $a_2$ via the
equation
\[R_2/a_2 \propto q^{1/3}(1+q)^{2/3},\]
which shows that the radius of the secondary star must increase if 
$q$ increases. This in turn means that the model line-width is greater than
the observed line-width and this overlap allows a greater range of 
secondary star geometries for which a satisfactory fit can be obtained.
In contrast, at low primary masses the allowed value of $q$ is decreased,
resulting in a low secondary star radius and hence a small model line-width.
At the lowest primary masses in figure~\ref{fig:entropyland}, 
the model line-widths become too narrow to fit the observed profile to the 
desired $\chi^2$, no matter what intensity pattern is mapped onto the 
secondary star's surface, explaining the abrupt end of the high entropy 
ridge on the lower left-hand portion of
figure~\ref{fig:entropyland}.

It is important to note that the maximum entropy algorithm we use requires 
that all input data is positive. As a result, it is necessary to invert 
absorption line profiles and, in the case of low signal-to-noise data, add a 
positive constant in order to ensure that there are no negative values in the 
continuum-subtracted spectra. If this positive constant is not added, and any 
negative points are either set to zero or ignored during the reconstruction, 
the fit will be positively biased in the line wings, resulting in a broader 
computed profile and hence larger stellar masses in the entropy landscape. 
It is a simple matter to account for this positive constant during the
reconstruction process by employing a virtual image element which contributes 
a single value to all data points, effectively cancelling out the constant. 
This virtual image element can also be used to correct for small errors in the
continuum subtraction, which might otherwise systematically affect the masses 
derived from the entropy landscape.

\section{Statistical errors}
\label{sec:statistical}

Although it is possible to propagate the statistical errors on the data
points through the iterative process in order to calculate the statistical
errors on each element in the map, the resulting error-bars are unreliable
due to the fact that noise in Roche tomograms is correlated. This
correlation arises from the projection of bumps in the profile along arcs of 
constant radial velocity on the secondary star and from the effects of 
blurring by the default map. Doppler imaging of single stars has
largely relied on consistency tests, such as comparing image reconstructions
using either subsets of observations (e.g. \pcite{barnes98}) or different 
spectral lines (e.g. \pcite{strassmeier00}), in order to address this issue. 
So far, the only attempt at a formal statistical error analysis in Doppler 
imaging has been undertaken using the Occamian approach \cite{berdyugina98}.

In the case of Roche tomography, one will undoubtedly be working with
lower signal-to-noise datasets and fewer spectral lines than
conventional Doppler imaging. For instance, the well-studied K-star
AB Dor has a mean visual magnitude of $\sim$ 7 \cite{cameron97}, whereas
the secondary stars in CVs never attain magnitudes of below $\sim$10 and
are usually several magnitudes fainter than this. Roche tomograms of 
CV secondaries are therefore not as strongly constrained as 
Doppler images of single stars by the input data. This makes the 
determination of statistical errors of prime importance in Roche tomography,
because only then can one test whether a surface feature is real or just a 
spurious artefact due to noise.

We have addressed the problem of statistical error determination in
Roche tomography using a Monte-Carlo style approach. 
Monte-Carlo techniques rely on the construction of a large (typically
hundreds, in the case of Roche tomography) sample of synthesized
datasets which have been effectively drawn from the same parent
population as the original dataset, i.e. as if the observations have
been repeated many hundreds of times. This large sample of
synthesized datasets is then used to create a large sample of Roche
tomograms, resulting in a probability distribution for each element in
the map. The main difficulty with this technique, aside from the
demands on  computer time, lies in the construction of the sample of
synthesized datasets. One approach (used by~\pcite{rutten94a} in
spectral eclipse mapping) is to
`jiggle' each data point about its observed value, by an
amount given by its error bar multiplied by a number output by a
Gaussian random-number generator with zero mean and unit
variance. This process adds noise to the data, however, which means
that the synthesized datasets are not being drawn from the same parent
population as the observed dataset -- noise is being added to a
dataset which has already had noise added to it during the measurement
process. In practice, this means that fitting the sample datasets to
the same level of $\chi^2$ as the original dataset is either
impossible (i.e. the iteration does not converge) or results in maps
dominated by noise, which grossly overestimates the true error on each 
element. We have verified that this is the case using simulations 
similar to those described in section~\ref{testing}.

Instead we have opted for the bootstrapping technique of \scite{efron79}; see 
\scite{efron93} for a general introduction to the method. We implemented the
bootstrap as follows. From our observed trailed spectrum containing {\em n}
data points we form our synthesized trailed spectrum by selecting, at random
and with replacement, {\em n} data values and placing these at their original
positions in the new synthesized trailed spectrum.
For points that are not selected we set the associated error bar to infinity
and thereby effectively omit these points from the fit. For points
that have been selected once or more than once (typically 37\% of the points)
we divide their error bars by the square root of the number of times they
were picked. The advantage of bootstrap
re-sampling over jiggling is that the data is not made noisier by the
process as only the errors bars on the data points are manipulated
(i.e. the data values remain unchanged). In so doing, the amount of noise 
present in the data is conserved and it is therefore possible to fit the 
synthesized trailed spectra to the same level of $\chi^2$ as the observed 
data, giving a much more reliable estimate of the statistical errors in the 
maps.

\subsection{Testing the bootstrap}
\label{testing}

\begin{figure}
\psfig{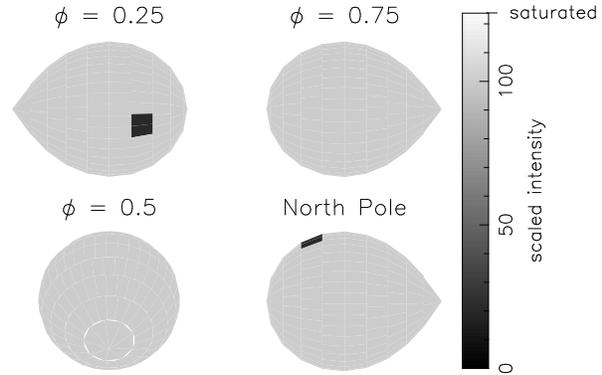}
\caption{The test image used to construct the noisy dataset described in
section~\ref{testing}}
\label{fig:stattest}
\end{figure}

In order to assess the viability of the bootstrap technique we ran a simple
test. A model with parameters M$_1$ = 1.0 M$_{\odot}$,
M$_2$ = 0.5 M$_{\odot}$, i = 60$^{\circ}$ and P = 2.78 hours was created
of uniform intensity distribution, with the exception of a single spot
covering 0.8\% of the total surface area on the trailing hemisphere (see
figure~\ref{fig:stattest}). From this model a synthetic trailed spectrum
was generated with 30 evenly spaced phase bins and 60 km\,s$^{-1}$
instrumental resolution. Each data point in the spectrum was then
`jiggled' by 10\% of the original data value in order to simulate noise.
A close fit to this dataset was then carried out using the correct
parameters.

\begin{figure}
\psfig{figure=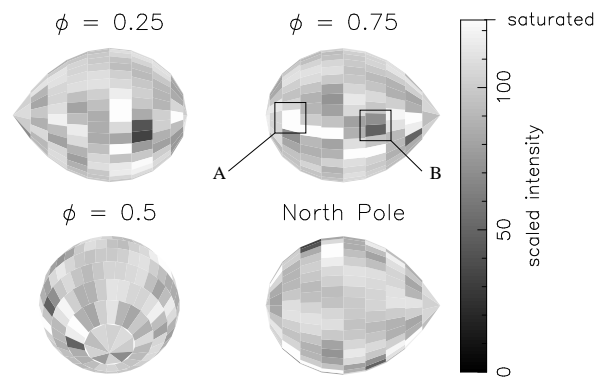,width=7.8cm,angle=-90}
\caption{The Roche tomography reconstruction of the test image in
figure~\ref{fig:stattest} using noisy data. Although the spot can be seen
at phase 0.25, many spurious artefacts due to noise are also present,
including features A and B.}
\label{fig:statrec}
\end{figure}

\begin{figure}
\psfig{figure=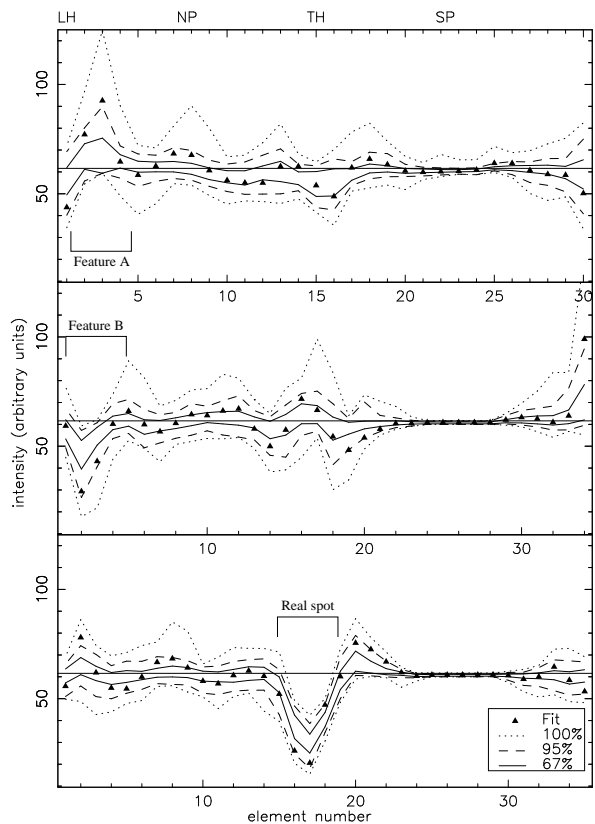,width=7.8cm}
\caption{Triangles: intensity values along vertical slices through A (top),
B (middle) and the real spot (bottom) in figure~\ref{fig:statrec}.
Curves: confidence intervals along vertical slices through A (top),
B (middle) and the real spot (bottom) in figure~\ref{fig:statrec}.
LH, NP, TH and SP at the top of the figure represent the positions of elements
at the leading
hemisphere, north pole, trailing hemisphere and south pole, respectively.}
\label{fig:slices}
\end{figure}

As can be seen, the reconstruction (figure~\ref{fig:statrec}) shows
many spurious artefacts due to noise. Fitting to a higher $\chi^2$,
whilst removing some of the lower-level noise artefacts, does not
remove the spurious spot features marked A and B in
figure~\ref{fig:statrec}. The triangular symbols in figure~\ref{fig:slices} 
show the intensities in each vertical slice through A, B and the real spot. 
It can be seen that, from this information alone, it is impossible to deduce
which spot features are real. 

\begin{figure}
\psfig{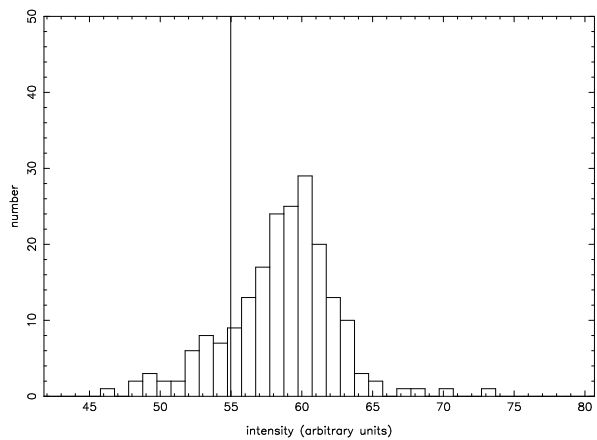}
\caption{Histogram of intensities obtained from 200 bootstrapped reconstructions for one element. The vertical line represents the intensity value
obtained in the original fit. The true value is 61.}
\label{fig:sample}
\end{figure}

The curves in figure~\ref{fig:slices} show
the confidence intervals found after reconstructing 200 bootstrap samples of 
the image. As the distribution of element
intensities found is often non-normal and not
centred on the intensity found in the original fit 
(see figure~\ref{fig:sample}), it is incorrect to
calculate a summary error statistic like the standard deviation. Instead
we take the mode of the distribution to represent the `most probable' value,
and define the 95\% confidence interval (for example) as the region which
encloses 95\% of the bootstrap values above and below the mode. It can be
seen that the confidence intervals in figure~\ref{fig:slices} widen
significantly around the noise features A and B, and encompass the true
intensity level (indicated by the horizontal line). The confidence intervals 
around the real spot feature, however, do not exhibit such a widening and 
clearly do not encompass the true intensity level of the non-spotted regions.
This test therefore confirms that the bootstrapping method works; it
has correctly identified the real spot feature and correctly identified 
features A and B as artefacts due to noise. In a similar test, statistical
errors determined by the method of `jiggling' failed to distinguish between
the real spot and features A and B.

There are two additional features of note in  figure~\ref{fig:slices}. 
First, it can be seen that in regions where the 
data do not constrain the intensity distribution (such as the southern 
hemisphere, which is never visible), the reconstruction always converges to 
the default map value, which is why the confidence intervals converge 
around `SP' in figure~\ref{fig:slices}. Second, the slices
through features A and B demonstrate how noise is correlated in the
Roche tomography reconstruction process, because elements surrounding
the peak intensity are all shifted in the same direction as the peak itself.

\section{Conclusions}
\label{sec:discussion}

We have shown that any feature on a Roche tomogram must be subjected
to two tests before its reality can be confirmed.  The first test is
to determine whether the feature is statistically significant and is
performed using a Monte-Carlo technique. The conventional method,
where hypothetical datasets are constructed by `jiggling' the data in
accordance with their error bars, was found to add noise and hence
grossly over-estimate the errors in the reconstruction. The solution
to this problem was found by using a bootstrap re-sampling algorithm
and, through the use of simulations, we have shown this technique to
correctly distinguish between real features and artefacts due to
noise.  The second test is to compare the feature with the appearance
of known artefacts of the technique. In general, we find that
systematic errors result in only three major artefacts in the Roche
tomograms: ring-like streaks, equatorial banding and blurring. 
The effects of these  artefacts can be minimised, however, by fine tuning the 
input parameters. A fine example of this is given by the entropy landscape 
technique, which also provides accurate component masses. If a surface feature
survives both of the above tests unscathed, it can be assumed to be real.

\section*{\sc Acknowledgements}

We would like to thank Chris Careless, Antonio Claret, Tom Marsh, 
Ren\'{e} Rutten and Tariq Shahbaz for their help in the work which
led to this paper. CAW is supported by a PPARC studentship. The
authors acknowledge the data analysis facilities at Sheffield
provided by the Starlink Project which is run by CCLRC on behalf of PPARC.

\bibliographystyle{mnras}
\bibliography{refs}

\end{document}